\def\etal{{\it et al.}}
\def\ie{{\it i.e.}}

\def\~{{$\tilde{\phantom{a}}$}}

\documentclass [12pt] {article}
\usepackage{epsfig}
\usepackage{color}

\def\thebibliography#1{\section{References}\markboth
 {REFERENCES}{REFERENCES}\list
 {[\arabic{enumi}]}{\settowidth\labelwidth{[#1]}\leftmargin\labelwidth
 \advance\leftmargin\labelsep
 \usecounter{enumi}}
 \def\newblock{\hskip .11em plus .33em minus -.07em}
 \sloppy
 \sfcode`\.=1000\relax}
\def\upcite#1{\raise6pt\hbox{\scriptsize
\cite{#1}}}
  \def\lsim{\mathrel {\vcenter {\baselineskip 0pt \kern 0pt
    \hbox{$<$} \kern 0pt \hbox{$\sim$} }}}
    \def\gsim{\mathrel {\vcenter {\baselineskip 0pt \kern 0pt
    \hbox{$>$} \kern 0pt \hbox{$\sim$} }}}

\def\hline{\noalign{\hrule \vskip2pt}}

\def\|{\ifmmode\Vert\else \char`\|\fi}
\ifx\oldzeta\undefined                          % hasn't been done yet, so 
  \let\oldzeta=\zeta                            % save old definiton
  \def\zzeta{{\raise 2pt\hbox{$\oldzeta$}}}     % make new definition
  \let\zeta=\zzeta                              % and attatch it
\fi

\ifx\oldchi\undefined                           % hasn't been done yet, so 
  \let\oldchi=\chi                              % save old definiton
  \def\cchi{{\raise 2pt\hbox{$\oldchi$}}}       % make new definition
  \let\chi=\cchi                                % and attatch it
\fi

\def\frac#1#2{{#1 \over #2}}

\def\half{\ifinner {\scriptstyle {1 \over 2}}
   \else {1 \over 2} \fi}

\def\simge{\mathrel{%
   \rlap{\raise 0.511ex \hbox{$>$}}{\lower 0.511ex \hbox{$\sim$}}}}
\def\simle{\mathrel{
   \rlap{\raise 0.511ex \hbox{$<$}}{\lower 0.511ex \hbox{$\sim$}}}}

\def\buildchar#1#2#3{{\null\!                   % \null, cancel space
   \mathop#1\limits^{#2}_{#3}                   % #1, #2 above, #3 below
   \!\null}}                                    % cancel space, \null
\def\overcirc#1{\buildchar{#1}{\circ}{}}

\def\slashchar#1{\setbox0=\hbox{$#1$}           % set a box for #1 
   \dimen0=\wd0                                 % and get its size
   \setbox1=\hbox{/} \dimen1=\wd1               % get size of /
   \ifdim\dimen0>\dimen1                        % #1 is bigger
      \rlap{\hbox to \dimen0{\hfil/\hfil}}      % so center / in box
      #1                                        % and print #1
   \else                                        % / is bigger
      \rlap{\hbox to \dimen1{\hfil$#1$\hfil}}   % so center #1
      /                                         % and print /
   \fi}                                         %

\def\subrightarrow#1{%                          % #1 under arrow
  \setbox0=\hbox{%                              % set a box
    $\displaystyle\mathop{}%                    % no mathop
    \limits_{#1}$}%                             % just limits
  \dimen0=\wd0%                                 % get width
  \advance \dimen0 by .5em%                     % add a bit
  \mathrel{%                                    % space like =
    \mathop{\hbox to \dimen0{\rightarrowfill}}% % arrow to width
       \limits_{#1}}}                           % text below

\def\overlay#1#2{\ifmmode%
\setbox0=\hbox{$#1$}%
\setbox1=\hbox to\wd0{\hss$#2$\hss}\else%
\setbox0=\hbox{#1}%
\setbox1=\hbox to\wd0{\hss#2\hss}\fi%
#1\hskip-\wd0\box1 }

\def\pmb#1{\leavevmode\setbox0=\hbox{#1}%
\kern-.02em\copy0\kern-\wd0
\kern.04em\copy0\kern-\wd0
\kern-.02em\raise.04em\box0 }

\def\vereq#1#2{\lower3pt\vbox{\baselineskip1.5pt \lineskip1.5pt
\ialign{$\m@th#1\hfill##\hfil$\crcr#2\crcr\sim\crcr}}}

\def\tensor#1{\protect\@ontopof{#1}{\leftrightarrow}{1.15}\mathord{\box2}}
\def\overstar#1{\protect\@ontopof{#1}{\ast}{1.15}\mathord{\box2}}
\def\overdots#1{\protect\@ontopof{#1}{\cdots}{1.0}\mathord{\box2}}
\def\overcirc#1{\protect\@ontopof{#1}{\circ}{1.2}\mathord{\box2}}
\def\loarrow#1{\protect\@ontopof{#1}{\leftarrow}{1.15}\mathord{\box2}}
\def\roarrow#1{\protect\@ontopof{#1}{\rightarrow}{1.15}\mathord{\box2}}

\def\@ontopof#1#2#3{%
{\mathchoice
{\@@ontopof{#1}{#2}{#3}\displaystyle\scriptstyle}%
{\@@ontopof{#1}{#2}{#3}\textstyle\scriptstyle}%
{\@@ontopof{#1}{#2}{#3}\scriptstyle\scriptscriptstyle}%
{\@@ontopof{#1}{#2}{#3}\scriptscriptstyle\scriptscriptstyle}%
}%
}

\def\@@ontopof#1#2#3#4#5{%
\setbox0=\hbox{$#4#1$}%
\setbox1=\hbox{$#5#2$}%
\setbox2=\hbox{}\ht2=\ht0 \dp2=\dp0 %
\ifdim\wd0>\wd1 %
\setbox1=\hbox to\wd0{\hss\box1\hss}%
\mathord{\rlap{\raise#3\ht0\box1}\box0}%
\else   %
\setbox1=\hbox to.9\wd1{\hss\box1\hss}%
\setbox0=\hbox to\wd1{\hss$#4\relax#1$\hss}%
\mathord{\rlap{\copy0}\raise#3\ht0\box1}%
\fi
}%

\def\lambdabar{\protect\@lambdabar}
\def\@lambdabar{%
\relax
\bgroup
\def\@tempa{\hbox{\raise.73\ht0
\hbox to0pt{\kern.25\wd0\vrule width.5\wd0
height.1pt depth.1pt\hss}\box0}}%
\mathchoice{\setbox0\hbox{$\displaystyle\lambda$}\@tempa}%
{\setbox0\hbox{$\textstyle\lambda$}\@tempa}%
{\setbox0\hbox{$\scriptstyle\lambda$}\@tempa}%
{\setbox0\hbox{$\scriptscriptstyle\lambda$}\@tempa}%
\egroup
}

\def\corresponds{{\lower.2ex\hbox{=}}{\rm\kern-.75em^\triangle}}
\def\succsim{\succ\kern-.9em_\sim\kern.3em}
\def\precsim{\prec\kern-1em_\sim\kern.3em}
\def\slantfrac#1#2{\kern1em^{#1}\kern-.3em/\kern-.1em_{#2}}

\begin{document}

\begin{center}
{\Large\bf A bounded source cannot emit a unipolar electromagnetic wave}
\\

\medskip

Kwang-Je Kim
\\
{\sl Argonne National Laboratory, Argonne, IL 60439}
\\
Kirk T.~McDonald
\\
{\sl Joseph Henry Laboratories,
Princeton University, Princeton, New Jersey 08544}
\\
Gennady V.~Stupakov
\\
{\sl Stanford Linear Accelerator Center, Stanford University, 
Stanford, CA 94309}
\\
Max S.~Zolotorev
\\
{\sl Center for Beam Physics, Lawrence Berkeley National Laboratory,
Berkeley, CA 94720}
\\
(May 1, 1999)
\end{center}

\section{Problem}

Show %, by considerations of energy conservation and Fourier analysis, 
that a bounded source cannot produce a unipolar electromagnetic pulse.

Equivalently, show that there are no three-dimensional electromagnetic
solitons in vacuum.
 
\section{Solution}

This problem was first discussed by Stokes over 150 years ago
\cite{Stokes,Rayleigh,Baker,Landau6}, who noted that
three-dimensional sound waves from a bounded source  cannot be unipolar. 
While his argument applies to electromagnetic waves as well, this is little
recognized in the literature.
One-dimensional electromagnetic (and sound) waves can be unipolar.  A 
plane wave can have any pulseform, but, strictly speaking, a plane wave
can only be generated by an unbounded source.  
A well-known pedagogic example given in sec.~II-20 of \cite{Feynman}
is based on this case. 

Hence, this old problem still has a new flavor.  Here, we offer two new 
solutions, followed by discussion.

\subsection{Via Conservation of Energy and Fourier Analysis}

If the source is bounded, it appears pointlike when viewed from a great
enough distance.  Then, energy conservation requires that the pulse
energy density fall off as $1/r^2$, for distance $r$ measured from some
characteristic point within the source.  Since the energy density
is proportional to the square of the the electromagnetic fields,
we have the well-known result that the radiation fields from a bounded
source fall off as $1/r$ far from the source.

This is in contrast to the static fields, which must
fall off at least as quickly at $1/r^2$, far from a bounded source.

Now, consider the possibility of a unipolar pulse, \ie, one for which
the electric field components $E_i({\bf r},t)$ have only one sign.
At a fixed point {\bf t}, the time integral of at least one component of
such a pulse would be nonzero:
\begin{equation}
\int E_i({\bf r},t)\, dt \neq 0.
\label{eq1}
\end{equation}
Then a Fourier analysis of this component,
\begin{equation}
E_i({\bf r},\omega) = \int E_i({\bf r},t) e^{i\omega t} dt,
\label{eq2}
\end{equation}
would have a nonzero value at zero frequency, 
$E_i({\bf r},\omega = 0) \neq 0$.
 
However, the quantity $E_i({\bf r},\omega = 0)$ would then be a static
solution to Maxwell's equation, and so must fall off like $1/r^2$.
This contradicts the hypothesis that $E_i({\bf r},t)$ represented an
electromagnetic pulse from a bounded source, which must fall off as
$1/r$.

Thus, a bounded source cannot emit a unipolar electromagnetic pulse.

\subsection{Via the Fields of an Accelerated Charge}

The electric field vector radiated by a charge is opposite to the transverse 
component of the acceleration.

In the case of a bounded source, the accelerations of the charges cannot 
impart a nonzero 
average velocity to any charge; otherwise the source would not remain bounded.  

Hence, any accelerations must include both positive and negative components 
such that their time integral vanishes.  

Therefore, the radiated electric field must also include both positive and 
negative components.

Again, A bounded source cannot emit a unipolar electromagnetic pulse.

A variant of the above argument using the Li\'enard-Wiechert fields has been
given by Bessonov \cite{Bessonov}, who also considers the case of a bipolar
pulse consisiting of a pair of well separated, opposite-sign unipolar pulses.

\subsection{Three-Dimensional Unipolar Radiation from an Unbounded Source}

In the case of two nonrelativistic, unbound charged particles that interact 
only via the Coulomb force $q_1 q_2/r$, the component of the acceleration of 
one of the
charges along the axis of its hyperbolic trajectory always has the same
sign.  Hence, the radiated electric field component along that axis is unipolar,
and a Fourier analysis of the field has a zero-frequency component (see sec.~70
of \cite{Landau}).  Thus, a three-dimensional unipolar pulse can be emitted by
a system whose motion is unbounded.

\subsection{Unipolarlike Pulses}

It is possible for a bounded system to produce an electromagnetic pulse that
consists almost
entirely of a single central pulse of one sign.  But according to
the argument above, this pulse must include long tails of the
opposite sign so that the time integral of the fields vanish at any
point far from the source.  This behavior has been observed in several
recent reports on subcycle electromagnetic pulses
\cite{Greene,You,Dormier,Bonvalet}.
 
% These tails were overlooked in a recent paper
%by Rau \etal\ \cite{Rau}.  They considered an electromagnetic pulse that
%is based on a so-called Bessel beam \cite{Durnin}, but an error in their
%eq.~(7) resulted in the factor $\exp[-k_0^2 \sigma^2 \rho^2 / 2 (1 + \rho^2)]$
%being written as $\exp[-k_0^2 \sigma^2 / 4 (1 + \rho^2)]$, which led them to
%believe that the pulse was unipolar.  A more realistic
%assessment of the use of Bessel beams for particle acceleration has been
%given by Hafizi \etal\ \cite{Hafizi}.

\section{Can the Far-Zone Radiation From a Bounded Source Transfer Energy to a
Charged Particle?}

The present considerations can be extended to comment on this topical 
question.  

If a unipolar electromagnetic pulse existed far from its bounded source,
the corresponding vector
potential {\bf A} would have different values at asymptotically early
and late
times.  Then, as argued by Lai \cite{Lai},
an interaction with a charge $e$ and mechanical momentum
{\bf P} that conserves the canonical momentum ${\bf P} + e{\bf A}/c$
could produce a net change in the magnitude of the mechanical
momentum, \ie, provided a transfer of energy between the particle and
the pulse.  Such far-zone unipolar particle acceleration is desirable,
but is not consistent with Maxwell's equations.

Near-zone particle acceleration by time-dependent electromagnetic
fields can be accomplished by passing a particle through a bounded
field region, such as an rf cavity, during a short interval when the
fields have only one sign.  While the electromagnetic fields are not
unipolar in this case, their interaction with the charged particle
is effectively so. 

A variant on such considerations is the fact that a bounded 
electrostatic field cannot exchange net energy with a charged particle
that begins and ends its history at large distances from the source.
So-called electrostatic particle accelerators all contain a 
nonelectrostatic component that can move the charge to a region of
nonzero electric potential and leave it there with effectively zero
electrical and mechanical energy.  Then, the charge can extract
energy from the field as it is expelled to large distances. 

Returning to the case of far-zone radiation,
we give a brief argument based the well-known relation for the time rate of
change of energy $U$ of a particle of charge $e$ and velocity {\bf v} in an 
electromagnetic field {\bf E} (see, for example, eq.~(17.7) of \cite{Landau}):
\begin{equation}
{dU \over dt} = e {\bf E} \cdot {\bf v}
\label{s3.1}
\end{equation}
This expression holds for particles of any velocity less than the speed of
light.  Of course, the magnetic field cannot change the particle's energy.

In the approximation that the particle's velocity is essentially
unchanged by its interaction with the electromagnetic field, we have
\begin{equation}
\Delta U = e {\bf v} \cdot \int {\bf E}\, dt.
\label{s3.2}
\end{equation}
To perform the integral, we can use Feynman's expression for the far-zone
radiated electric field of an accelerated charge (sec.~I-34 of \cite{Feynman}):
\begin{equation}
{\bf E}_{\rm rad} = {e \over c^2} {d^2 \hat{\bf n} \over dt^2}\, ,
\label{s3.3}
\end{equation}
in Gaussian units,
where $\hat{\bf n}$ is a unit vector from the retarded position of the source
charge to the observer.  Then, $\int {\bf E}_{\rm rad}\, dt$ is the difference
between $d\hat{\bf n}/dt$ at early and late times.  Since $d\hat{\bf n}/dt$ is
the angular velocity of the relative motion between the source and charge, 
this vanishes at both early and late times as the moving charge is then
arbitrarily far from the bounded source.
Hence, in the constant-velocity approximation,
the far-zone fields from a bounded source cannot transfer energy to a 
particle.

While the constant-velocity
approximation is not necessarily good for a particle whose
initial velocity is nonrelativistic, it is an excellent approximation for
a relativistic particle.  It is possible for the far-zone fields of
a bounded source (in vacuum and far from any other material) to transfer
energy to a nonrelativistic charge, as recently observed \cite{Malka},
but this energy transfer becomes
increasingly inefficient as the particle's velocity approaches that of light.
%So-called vacuum laser acceleration could be effective  mainly in the 
%nonrelativistic
%regime, and is unlikely to compete with established techniques for near-zone
%acceleration of relativistic particles.

The above considerations have ignored energy transfers due to scattering, which
can be significant if the energy of a photon of the electromagnetic wave is
large compared to the total energy of the charged particle in the center of
mass frame.  In this case, a quantum description is more appropriate.
In the regime in which the energy transfer from a single scattering process
is small, the classical idea of a ``radiation pressure'' associated with
the Poynting vector ${\bf S} = (c/4\pi) {\bf E} \times {\bf H}$
can be formalized
by including the radiation reaction in the equation of motion of the
charged particle.  See, for example, eq.~(75.10) of \cite{Landau}.
However, this effect is always very small in the classical regime.

\end{document}